\def\be{\begin{equation}}
\def\ee{\end{equation}}
\def\ba{\begin{array}}

\def\ea{\end{array}}

\documentclass[12pt]{article}
\usepackage{amsfonts,amssymb,amsmath,graphicx}
\newtheorem{theorem}{Theorem}

\newtheorem{lemma}{Lemma}
\begin{document}
\parskip=3pt
\parindent=18pt
\baselineskip=20pt
\setcounter{page}{1}
\centerline{\large\bf Tighter weighted polygamy inequalities of multipartite entanglement}
\centerline{\large\bf in arbitrary-dimensional quantum systems}
\vspace{6ex}
\centerline{{\sf Bin Chen,$^{\star}$}
\footnote{\sf Corresponding author: chenbin5134@163.com}
~~~ {\sf Long-Mei Yang$^{\natural}$}
~~~ {\sf Shao-Ming Fei$^{\natural,\sharp}$}
~~~ {\sf Zhi-Xi Wang$^{\natural}$}
}
\vspace{4ex}
\centerline
{\it $^\star$ College of Mathematical Science, Tianjin Normal University, Tianjin 300387, China}\par
\centerline
{\it $^\natural$ School of Mathematical Sciences, Capital Normal University, Beijing 100048, China}\par
\centerline
{\it $^\sharp$ Max-Planck-Institute for Mathematics in the Sciences, 04103 Leipzig, Germany}\par
\vspace{6.5ex}
\parindent=18pt
\parskip=5pt
\begin{center}
\begin{minipage}{5in}
\vspace{3ex}
\centerline{\large Abstract}
\vspace{4ex}
We investigate polygamy relations of multipartite entanglement in arbitrary-dimensional quantum systems.
By improving an inequality and using the $\beta$th ($0\leq\beta\leq1$) power of entanglement of assistance, we provide a new class of weighted polygamy
inequalities of multipartite entanglement in arbitrary-dimensional quantum systems.
We show that these new polygamy relations are tighter than the ones given in [Phys. Rev. A \textbf{97}, 042332 (2018)].
\end{minipage}
\end{center}

\newpage

\section{Introduction}

Monogamy of entanglement (MOE) is one of the hot issues in the study of quantum information theory in recent years.
Being an intriguing feature of quantum entanglement, it is tightly related to many quantum information and communication processing tasks such as the security
proof in quantum cryptographic scheme \cite{CHB}.
Mathematically, MOE can be characterized by the following inequality:
\begin{equation}
E(\rho_{A|BC})\geq E(\rho_{A|B})+E(\rho_{A|C}),
\end{equation}
where $\rho_{ABC}$ is a tripartite quantum state with its reduced density matrices $\rho_{AB}={\rm tr}_C(\rho_{ABC})$ and $\rho_{AC}={\rm tr}_B(\rho_{ABC})$,
$E(\cdot)$ is a bipartite entanglement measure.
The first monogamy inequality was proposed for three-qubit systems by use of tangle as the bipartite entanglement measure \cite{VC},
and later was generalized to the case of multiqubit quantum systems \cite{TJO} as well as some cases of higher-dimensional quantum systems using various bipartite entanglement measures \cite{JSK1,JSK2,JSK3,JSK4}.

Besides MOE, polygamy of entanglement (POE) has also attracted much attention, due to its dually monogamous property in multipartite quantum systems.
POE is also mathematically described by the following inequality:
\begin{equation}
E_{a}(\rho_{A|BC})\leq E_{a}(\rho_{A|B})+E_{a}(\rho_{A|C}),
\end{equation}
where $E_{a}(\cdot)$ is the assisted entanglement \cite{Gour05}.
The first polygamy inequality was established in three-qubit systems by use of tangle of assistance \cite{Gour05}.
It was later generalized to multiqubit systems by using various assisted entanglements \cite{JSK3,JSK4,Gour07}.
For the case of arbitrary-dimensional quantum systems, general polygamy inequalities of multipartite entanglement is also proposed in Ref. \cite{Buscemi,JSK5,JSK6} by using entanglement of assistance.

In Ref. \cite{Gour05}, Gour \emph{et al.} proposed the first polygamy relation in three-qubit systems.
For a three-qubit pure state $|\psi\rangle_{ABC}$, the following polygamy inequality holds \cite{Gour05}:
\begin{equation}\label{ft}
\tau(|\psi\rangle_{A|BC})\leq\tau_{a}(\rho_{AB})+\tau_{a}(\rho_{AC}),
\end{equation}
where $\tau(|\psi\rangle_{A|BC})$ is the tangle of the pure state $|\psi\rangle_{ABC}$ under partition $A$ and $BC$, and $\tau_{a}(\rho_{AB})=\max\sum_{i}p_{i}\tau(|\psi_{i}\rangle_{AB})$ is the tangle of assistance of $\rho_{AB}=\mathrm{Tr}_{C}|\psi\rangle_{ABC}\langle\psi|$,
with the maximum taken over all possible pure-state decompositions of $\rho_{AB}=\sum_{i}p_{i}|\psi_{i}\rangle_{AB}\langle\psi_{i}|$.
After that, inequality (\ref{ft}) was generalized to the case of arbitrary $N$-qubit quantum systems \cite{Gour07}:
\begin{equation}\label{ftn}
\tau_{a}(\rho_{A_{1}|A_{2}\cdots A_{N}})\leq\sum_{i=2}^{N}\tau_{a}(\rho_{A_{1}A_{i}}),
\end{equation}
where $\rho_{A_{1}A_{2}\cdots A_{N}}$ is an $N$-qubit mixed state, and $\rho_{A_{1}A_{i}}$ are reduced density matrices, $i=2,\ldots,n$.

For the case of higher-dimensional quantum systems, the
von Neumann entropy is shown to be a good measure to establish polygamy inequalities of multipartite entanglement in arbitrary-dimensional systems \cite{Buscemi}.
For a tripartite pure state $|\psi\rangle_{ABC}$, one has \cite{Buscemi}
$E(|\psi\rangle_{A|BC})\leq E_{a}(\rho_{AB})+E_{a}(\rho_{AC})$,
where $E(|\psi\rangle_{A|BC})=S(\rho_{A})=-\mathrm{Tr}\rho_{A}\ln\rho_{A}$ is the von Neumann entropy of entanglement between $A$ and $BC$,
and $E_{a}(\rho_{AB})$ is the entanglement of assistance (EOA) of $\rho_{AB}$ defined by \cite{OC}, $E_{a}(\rho_{AB})=\max\sum_{i}p_{i}E(|\psi_{i}\rangle_{AB})$,
where the maximization is taken over all possible pure state decompositions of $\rho_{AB}=\sum_{i}p_{i}|\psi_{i}\rangle_{AB}\langle\psi_{i}|$.
Furthermore, for arbitrary dimensional quantum states $\rho_{A_{1}A_{2}\cdots A_{n}}$, a general polygamy inequality of multipartite quantum entanglement was also established \cite{JSK5},
\begin{equation}\label{JSK5}
E_{a}(\rho_{A_{1}|A_{2}\cdots A_{n}})\leq \sum_{i=2}^{n}E_{a}(\rho_{A_{1}|A_{i}}).
\end{equation}

In Ref. \cite{Weighted}, Kim proposed a tight polygamy inequality of multipartite entanglement in arbitrary-dimensional quantum systems.
Using the $\beta$th power of entanglement of assistance for $0\leq\beta\leq1$, and the Hamming weight of the binary vector related with the distribution of subsystems, a class of weighted polygamy inequalities of multipartite entanglement in arbitrary-dimensional quantum systems is derived.
These inequalities are shown to be tighter than the previous ones for some class of quantum states.
However, we claim that these results can be further improved by refining an inequality.
In this paper, we provide a new class of weighted polygamy inequalities of multipartite entanglement in arbitrary-dimensional quantum systems.
We show that our new polygamy inequalities are tighter than the one given in \cite{Weighted}.

\section{Tighter weighted polygamy inequalities of multipartite entanglement}

In this section, we establish strengthened weighted polygamy inequalities for multipartite entanglement.
Suppose that $j$ is a non-negative integer with its binary expansion $j=\sum_{i=0}^{n-1}j_{i}2^i$, where $\log_{2}j<n$ and $j_i\in\{0,1\}$ for $i=0,1,\ldots,n-1$.
The binary vector of $j$ is defined as $\vec{j}=(j_0,j_1,\ldots,j_{n-1})$, and the Hamming weight of $\vec{j}$, denoted as $\omega_H(\vec{j})$, is defined as the number of $1$'s in its coordinates \cite{MAN}.
By using the $\beta$th power of EOA and the Hamming weight of the binary vector related with the distribution of subsystems, the author in Ref. \cite{Weighted} provided the following weighted polygamy inequality of multipartite entanglement in arbitrary dimensional quantum systems:
\begin{equation}\label{kim1}
[E_{a}(\rho_{A_{1}|B_{0}B_{1}\cdots B_{N-1}})]^{\beta}\leq \sum_{j=0}^{N-1}\beta^{\omega_{H}(\vec{j})}[E_{a}(\rho_{A|B_{j}})]^{\beta}.
\end{equation}
In the following we show that this inequality can be further refined.
We first introduce the Lemma below \cite{Yang}.

\begin{lemma}
Suppose $k$ is a real number satisfying $0< k\leq1$, then for any $0\leq x\leq k$,
we have
\begin{equation}
(1+x)^\beta\leq1+\frac{(1+k)^\beta-1}{k^\beta}x^\beta,
\end{equation}
for $0\leq\beta\leq1$.
\end{lemma}

We have the following Theorem.

\begin{theorem}\label{thm1}
Suppose that $0\leq\beta\leq1$. For any $(N+1)$-partite quantum state $\rho_{AB_{0}B_{1}\cdots B_{N-1}}$, if the subsystems $B_{0},B_{1},\cdots,B_{N-1}$ satisfy the following condition,
\begin{equation}\label{cdt1}
E_{a}(\rho_{A|B_{0}})\geq\frac{1}{k}E_{a}(\rho_{A|B_{1}})\geq\frac{1}{k^{2}}E_{a}(\rho_{A|B_{2}})\geq\cdots\geq\frac{1}{k^{N-1}}E_{a}(\rho_{A|B_{N-1}})\geq0,
\end{equation}
where $0<k\leq1$, then
\begin{equation}\label{eqt1}
[E_{a}(\rho_{A_{1}|B_{0}B_{1}\cdots B_{N-1}})]^{\beta}\leq \sum_{j=0}^{N-1}\left[\frac{(1+k)^\beta-1}{k^\beta}\right]^{\omega_{H}(\vec{j})}[E_{a}(\rho_{A|B_{j}})]^{\beta}.
\end{equation}
\end{theorem}

\emph{Proof}. We prove the Theorem by an approach similar to the one in \cite{Weighted}.
We denote $E:=E_{a}(\rho_{A_{1}|B_{0}B_{1}\cdots B_{N-1}}), E_{j}=E_{a}(\rho_{A|B_{j}}), j=0,1,\cdots,N-1$, for simplicity.
Since it has been proved that $E^{\beta}\leq(\sum_{j=0}^{N-1}E_{j})^{\beta}$ \cite{JSK5}, we only need to show that
\begin{equation}\label{eqt11}
\left(\sum_{j=0}^{N-1}E_{j}\right)^{\beta}\leq\sum_{j=0}^{N-1}\left[\frac{(1+k)^\beta-1}{k^\beta}\right]^{\omega_{H}(\vec{j})}E_{j}^{\beta}.
\end{equation}

We first prove the inequality (\ref{eqt11}) for the case of $N=2^{n}$, and use the mathematical induction on $n$.
For $n=1$, using Lemma 1, one can easily get
\begin{equation}
\begin{split}
(E_{0}+E_{1})^{\beta}&=E_{0}^{\beta}\left(1+\frac{E_{1}}{E_{0}}\right)^{\beta}\\
&\leq E_{0}^{\beta}\left[1+\frac{(1+k)^\beta-1}{k^\beta}\left(\frac{E_{1}}{E_{0}}\right)^{\beta}\right]\\
&=E_{0}^{\beta}+\frac{(1+k)^\beta-1}{k^\beta}E_{1}^{\beta}.
\end{split}
\end{equation}
Thus the inequality (\ref{eqt11}) holds when $N=2$, i.e., $n=1$.

Assume that the inequality (\ref{eqt11}) holds for $N=2^{n-1}$.
We now consider the case of $N=2^{n}$.
From the conditions of the Theorem, we first have the following fact:
\begin{equation}
E_{2^{n-1}}\leq k^{2^{n-1}}E_{0}, E_{2^{n-1}+1}\leq k^{2^{n-1}}E_{1}, \cdots, E_{2^{n}-1}\leq k^{2^{n-1}}E_{2^{n-1}-1}.
\end{equation}
Summing up the above inequalities, we get
\begin{equation}
\begin{split}
E_{2^{n-1}}+E_{2^{n-1}+1}+\cdots+E_{2^{n}-1} &\leq k^{2^{n-1}}(E_{0}+E_{1}+\cdots+E_{2^{n-1}-1})\\
&\leq k(E_{0}+E_{1}+\cdots+E_{2^{n-1}-1}).
\end{split}
\end{equation}
Again using Lemma 1, we obtain
\begin{equation}
\begin{split}
\left(\sum_{j=0}^{2^{n}-1}E_{j}\right)^{\beta}&=\left(\sum_{j=0}^{2^{n-1}-1}E_{j}\right)^{\beta}\left(1+\frac{\sum_{j=2^{n-1}}^{2^{n}-1}E_{j}}{\sum_{j=0}^{2^{n-1}-1}E_{j}}\right)^{\beta}\\
&\leq\left(\sum_{j=0}^{2^{n-1}-1}E_{j}\right)^{\beta}\left[1+\frac{(1+k)^\beta-1}{k^\beta}\right]\left(\frac{\sum_{j=2^{n-1}}^{2^{n}-1}E_{j}}{\sum_{j=0}^{2^{n-1}-1}E_{j}}\right)^{\beta}\\
&=\left(\sum_{j=0}^{2^{n-1}-1}E_{j}\right)^{\beta}+\frac{(1+k)^\beta-1}{k^\beta}\left(\sum_{j=2^{n-1}}^{2^{n}-1}E_{j}\right)^{\beta}.
\end{split}
\end{equation}

From the induction hypothesis, we have
\begin{equation}
\left(\sum_{j=0}^{2^{n-1}-1}E_{j}\right)^{\beta}\leq \sum_{j=0}^{2^{n-1}-1}\left[\frac{(1+k)^\beta-1}{k^\beta}\right]^{\omega_{H}(\vec{j})}E_{j}^{\beta}.
\end{equation}
On the other hand, it is obvious that
\begin{equation}
\left(\sum_{j=2^{n-1}}^{2^{n}-1}E_{j}\right)^{\beta}\leq\sum_{j=2^{n-1}}^{2^{n}-1}\left[\frac{(1+k)^\beta-1}{k^\beta}\right]^{\omega_{H}(\vec{j})-1}E_{j}^{\beta}.
\end{equation}
Hence we have
\begin{equation}
\left(\sum_{j=0}^{2^{n}-1}E_{j}\right)^{\beta}\leq\sum_{j=0}^{2^{n}-1}\left[\frac{(1+k)^\beta-1}{k^\beta}\right]^{\omega_{H}(\vec{j})}E_{j}^{\beta}.
\end{equation}
This indicates that the inequality (\ref{eqt11}) holds for $N=2^{n}$.

Let $N$ be an arbitrary positive integer. Then there exists $n$ such that $0<N\leq2^{n}$.
Consider a $(2^{n}+1)$-partite quantum state $\gamma_{A|B_{0}B_{1}\cdots B_{2^{n}-1}}=\rho_{A|B_{0}B_{1}\cdots B_{N-1}}\otimes\sigma_{B_{N}\cdots B_{2^{n}-1}}$.
As just proved above we have that
\begin{equation}
[E_{a}(\gamma_{A|B_{0}B_{1}\cdots B_{2^{n}-1}})]^{\beta}\leq\sum_{j=0}^{2^{n}-1}\left[\frac{(1+k)^\beta-1}{k^\beta}\right]^{\omega_{H}(\vec{j})}[E_{a}(\gamma_{A|B_{j}})]^{\beta},
\end{equation}
where $\gamma_{A|B_{j}}$ are reduced density matrices of $\gamma_{A|B_{0}B_{1}\cdots B_{2^{n}-1}}$, $j=0,1,\cdots,2^{n}-1$.
Taking into account that $E_{a}(\gamma_{A|B_{0}B_{1}\cdots B_{2^{n}-1}})=E_{a}(\rho_{A|B_{0}B_{1}\cdots B_{N-1}})$
and $E_{a}(\gamma_{A|B_{j}})=0$, for $j=N,\cdots,2^{n}-1$, we have
\begin{equation}
\begin{split}
[E_{a}(\rho_{A|B_{0}B_{1}\cdots B_{N-1}})]^{\beta}&=[E_{a}(\gamma_{A|B_{0}B_{1}\cdots B_{2^{n}-1}})]^{\beta}\\
&\leq\sum_{j=0}^{2^{n}-1}\left[\frac{(1+k)^\beta-1}{k^\beta}\right]^{\omega_{H}(\vec{j})}[E_{a}(\gamma_{A|B_{j}})]^{\beta}\\
&=\sum_{j=0}^{N-1}\left[\frac{(1+k)^\beta-1}{k^\beta}\right]^{\omega_{H}(\vec{j})}[E_{a}(\rho_{A|B_{j}})]^{\beta},
\end{split}
\end{equation}
where the last equality holds since $\gamma_{A|B_{j}}=\rho_{A|B_{j}}$, for $j=0,\cdots,N-1$.
This completes the proof. $\Box$

Note that when $k=1$, the inequality (\ref{eqt1}) becomes
\begin{equation}\label{z}
[E_{a}(\rho_{A_{1}|B_{0}B_{1}\cdots B_{N-1}})]^{\beta}\leq \sum_{j=0}^{N-1}(2^{\beta}-1)^{\omega_{H}(\vec{j})}[E_{a}(\rho_{A|B_{j}})]^{\beta}.
\end{equation}
Since $2^{\beta}-1\leq\beta$ for $0\leq\beta\leq1$, this new inequality is tighter than inequality (\ref{kim1}).
To be more intuitive, let us consider the three qubit $W$ state, $|W\rangle_{ABC}=\frac{1}{\sqrt{3}}(|100\rangle+|010\rangle+|001\rangle)$.
One has $E_{a}(\rho_{A|BC})=S(\rho_{A})=\log_{2}3-\frac{2}{3}, E_{a}(\rho_{A|B})=E_{a}(\rho_{A|C})=\frac{2}{3}$ \cite{Sahoo}.
Then inequality (\ref{z}) yields that $\sqrt{E_{a}(\rho_{A|B})}+(\sqrt{2}-1)\sqrt{E_{a}(\rho_{A|C})}-\sqrt{E_{a}(\rho_{A|BC})}\approx0.196$ when $\beta=\frac{1}{2}$, while $\sqrt{E_{a}(\rho_{A|B})}+\frac{1}{2}\sqrt{E_{a}(\rho_{A|C})}-\sqrt{E_{a}(\rho_{A|BC})}\approx0.272$ from inequality (\ref{kim1}), which shows that our inequality is tighter than the one in (\ref{kim1}). See also Fig.1.

\begin{figure}
\centering
\includegraphics[width=7cm]{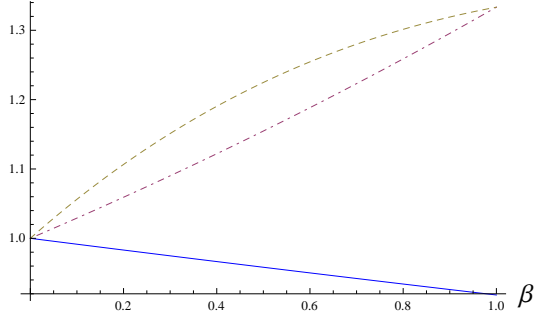}
\caption{The $y$ axis is the EOA of $|W\rangle_{ABC}$ and its upper bound, which are functions of $\beta$. The blue solid line is the EOA of $|W\rangle_{ABC}$, 
the dot-dashed line is the upper bound from our result, and the dashed line is the upper bound from the result in \cite{Weighted}.}
\end{figure}

In Ref. \cite{Weighted}, the author derived another weighted polygamy inequality which is tighter than inequality (\ref{kim1}):
\begin{equation}
[E_{a}(\rho_{A_{1}|B_{0}B_{1}\cdots B_{N-1}})]^{\beta}\leq \sum_{j=0}^{N-1}\beta^{j}[E_{a}(\rho_{A|B_{j}})]^{\beta},
\end{equation}
conditioned that $E_{a}(\rho_{A|B_{i}})\geq\sum_{j=i+1}^{N-1}E_{a}(\rho_{A|B_{j}})$, for $i=0,1,\cdots,N-2$.
Here we can also establish a strengthened type of the weighted polygamy inequality given in Theorem 1.

\begin{theorem}
For any $(N+1)$-partite quantum state $\rho_{AB_{0}B_{1}\cdots B_{N-1}}$ and $0\leq\beta\leq1$, if the following condition is satisfied
\begin{equation}
E_{a}(\rho_{A|B_{i}})\geq\frac{1}{k}\sum_{j=i+1}^{N-1}E_{a}(\rho_{A|B_{j}})
\end{equation}
for $i=0,1,\cdots,N-2$, where $0<k\leq1$, then
\begin{equation}
[E_{a}(\rho_{A_{1}|B_{0}B_{1}\cdots B_{N-1}})]^{\beta}\leq \sum_{j=0}^{N-1}\left[\frac{(1+k)^\beta-1}{k^\beta}\right]^{j}[E_{a}(\rho_{A|B_{j}})]^{\beta}.
\end{equation}
\end{theorem}

\emph{Proof}. We follow the proof by Kim \cite{Weighted}.
From inequality (\ref{JSK5}), it is sufficient to show that
\begin{equation}\label{yz}
\left(\sum_{j=0}^{N-1}E_{a}(\rho_{A|B_{j}})\right)^{\beta}\leq \sum_{j=0}^{N-1}\left[\frac{(1+k)^\beta-1}{k^\beta}\right]^{j}[E_{a}(\rho_{A|B_{j}})]^{\beta}.
\end{equation}
We use the mathematical induction on $N$. It is obvious that inequality (\ref{yz}) holds when $N=2$. Assume that it also holds for any positive integer less than $N$. Consider the state $\rho_{AB_{0}B_{1}\cdots B_{N-1}}$.
Since $0\leq\frac{\sum_{j=1}^{N-1}E_{a}(\rho_{A|B_{j}})}{E_{a}(\rho_{A|B_{0}})}\leq k$, we have
\begin{equation}\label{d}
\begin{split}
\left(\sum_{j=0}^{N-1}E_{a}(\rho_{A|B_{j}})\right)^{\beta}&=[E_{a}(\rho_{A|B_{0}})]^{\beta}\left(1+\frac{\sum_{j=1}^{N-1}E_{a}(\rho_{A|B_{j}})}{E_{a}(\rho_{A|B_{0}})}\right)^{\beta}\\
&\leq[E_{a}(\rho_{A|B_{0}})]^{\beta}\left[1+\frac{(1+k)^\beta-1}{k^\beta}\left(\frac{\sum_{j=1}^{N-1}E_{a}(\rho_{A|B_{j}})}{E_{a}(\rho_{A|B_{0}})}\right)^{\beta}\right]\\
&=[E_{a}(\rho_{A|B_{0}})]^{\beta}+\frac{(1+k)^\beta-1}{k^\beta}\left(\sum_{j=1}^{N-1}E_{a}(\rho_{A|B_{j}})\right)^{\beta},
\end{split}
\end{equation}
where the inequality is due to Lemma 1.
On the other hand, from the induction hypothesis we get
\begin{equation}\label{dd}
\left(\sum_{j=1}^{N-1}E_{a}(\rho_{A|B_{j}})\right)^{\beta}\leq\sum_{j=1}^{N-1}\left[\frac{(1+k)^\beta-1}{k^\beta}\right]^{j-1}[E_{a}(\rho_{A|B_{j}})]^{\beta}.
\end{equation}
Taking in to account (\ref{dd}) and the last equality of (\ref{d}), we complete the proof. $\Box$

\section{Conclusion}
We have studied polygamy relations of multipartite entanglement in arbitrary-dimensional quantum systems.
Inspired by the previous work in \cite{Weighted}, we have proposed a new class of weighted polygamy inequalities for multipartite entanglement in arbitrary-dimensional quantum systems by using the $\beta$th power of entanglement of assistance for $0\leq\beta\leq1$.
We have proved that these new inequalities are tighter than the ones given in \cite{Weighted}.
Our results may shed new light on the research of polygamy constraints of multipartite entanglement
and provide finer characterization of the entanglement distributions.

\vspace{2.5ex}
\noindent{\bf Acknowledgments}\, \,
This work is supported by the National Natural Science Foundation of China under Grant Nos. 11805143 and 11675113, and NSF of Beijing under No. KZ201810028042.

\end{document}